\documentclass[aps,prl,twocolumn,amsmath,amssymb,amsfonts]{revtex4-2}
\usepackage{dcolumn}
\usepackage{bm}
\usepackage{graphicx} 
\usepackage{epstopdf}

\usepackage{xcolor}
\newcommand{\be}{\begin{equation}}
\newcommand{\ee}{\end{equation}}
\newcommand{\bea}{\begin{eqnarray}}
\newcommand{\eea}{\end{eqnarray}}
\newcommand{\rIm}{\mathrm{Im}}
\newcommand{\rRe}{\mathrm{Re}}
\newcommand{\re}{\mathrm{e}}            
\newcommand{\ri}{\mathrm{i}}
\usepackage[titletoc,title]{appendix}
\usepackage[normalem]{ulem} 

\newcommand{\red}[1]{{\color{red}#1}}

\begin{document}

\title{Quartz Crystal Microbalance frequency response to discrete adsorbates in liquids}

\author{Alexander M. Leshansky$^1$}\email{lisha@technion.ac.il}
\author{Itzhak Fouxon$^{1}$}
\author{Boris Y. Rubinstein$^2$} 
\affiliation{$^1$Department of Chemical Engineering, Technion, Haifa 32000, Israel}
\affiliation{$^2$Stowers Institute for Medical Research, 1000 E 50th st., Kansas City, MO 64110, USA}

\begin{abstract}
Quartz Crystal Microbalance with Dissipation monitoring (QCM-D) has become a major tool in the gravimetric analysis of nanometric 
objects, such as proteins, viruses, liposomes and inorganic particles from the solution. While in vacuum extremely accurate measurements are possible, in a liquid phase the quantitative analysis is intricate due to the complex interplay of hydrodynamic and adhesion forces, varying with the physicochemical properties of adsorbate and the quartz resonator surfaces. In the present paper we dissect the role of hydrodynamics for the analytically tractable scenario of a \emph{stiff} contact, whereas the adsorbed \emph{rigid} particles oscillate with the resonator as a whole without rotation. Under the assumption of the low surface coverage, we theoretically study the \emph{excess} shear force exerted on the resonator due to presence of a single particle. The excess shear force has two contributions: (i) the \emph{fluid-mediated} force due to flow disturbance created by the particle and (ii) the viscous force exerted on the particle by the fluid and transmitted to the resonator \emph{via contact}. We found that for small adsorbates, there is mutual cancellation of the above contributions to the excess shear force at the leading order approximation, reducing the overall effect of the hydrodynamics to the order-of-magnitude of the inertial force. However, accurate numerical solution shows that for small particles the viscous force dominates over the inertia force, rendering the standard Sauerbrey model inapplicable. These findings indicate that the accurate account of hydrodynamics in the analysis of QCM-D response is critical. The resulting dimensionless frequency and dissipation shifts and the corresponding acoustic ratio computed numerically, showing a fair agreement with previously published experimental results at low oscillation frequencies. 

\end{abstract}

\maketitle

\noindent \textit{Introduction.}
Quartz crystal microbalance (QCM) technique \cite{review2015,johan21} relies on the fact that matter adsorbed on the surface of the fast oscillating crystal, changes the frequency of the oscillations. In vacuum, the shift in the resonant frequency of the crystal is linearly proportional to the mass of the adsorbed film via the seminal Sauerbrey equation \cite{sauerbrey59}, allowing extremely accurate measurements down to nanograms \cite{johan21}. The quantitative interpretation of the QCM-D measurement in liquids \cite{nomura80, qcm-liquid} (where ``D" stands for dissipation monitoring via measuring the decay rate of the oscillations) is also well-established for planar (including viscoelastic \cite{johan21}) adsorbed films.  However, interpreting the QCM-D measurements due to \emph{discrete} adsorbates  (such as, e.g., nanoparticles, liposomes, viruses, proteins, etc.) in liquids remains a challenge mainly due to the interplay of complex hydrodynamics, which has not yet been yet fully resolved and \emph{a priori} unknown viscoelastic contact dynamics, which depends on physicochemical properties of the surfaces (i.e., the adsorbate and the resonator) \cite{johan21}.  

The impedance $\mathcal{Z}$ probed by the QCM-D is the ratio $\overline{\sigma}/v_c$, where $\overline{\sigma}$ is the area-averaged tangential stress (i.e. the net shear force $\mathcal{F}$ exerted on the surface of the oscillating quartz resonator divided by its surface area) and $v_c$ is the velocity of the crystal oscillations. Here $\mathcal{F}$ and $v_c$ and, therefore, $\mathcal{Z}$ are all complex quantities characterized by the amplitude and phase. In the framework of the \emph{small load approximation} the shift in oscillation frequency, $\Delta f$, and in half-bandwidth, $\Delta \Gamma$ (related to a dissipation factor $\Delta \mathcal D$), are linearly proportional to the impedance,  $\Delta f-\ri \Delta\Gamma=\ri f_0 \mathcal{Z}/(\pi \mathcal{Z}_q)$, where $f_0$ stands for the fundamental oscillation frequency (typically 5~MHz) and the resonator's shear-wave impedance $\mathcal{Z}_q$ is a known quantity \cite{review2015,johan21}. The small load approximation holds given that $|\Delta f| \ll f_0$. In liquids, in contrast to vacuum where the adsorbed particles only alter the mass (i.e., solid inertia) of the resonator contributing to the frequency shift, $\Delta f$, according to the Sauerbrey equation, the adsorbed particles modify the viscous \emph{shear force} exerted onto the resonator, contributing to the shifts in the resonant frequency, $\Delta f$, and the bandwidth, $\Delta\Gamma$ (absent in vacuum). 

\begin{figure}[t]
\includegraphics[width=0.80\columnwidth]{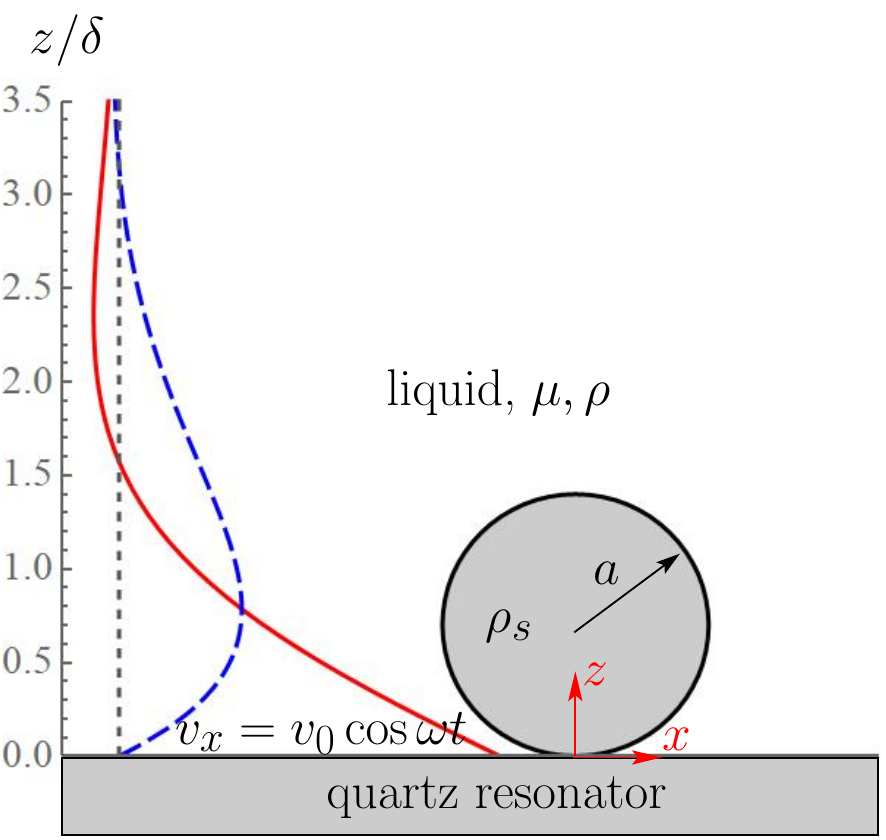}
\caption{Schematic illustration of the problem. A spherical particle of radius $a$ immersed in an incompressible viscous liquid of density $\rho$ and viscosity $\eta$ is rigidly attached to an infinite horizontal plane at $z\!=\!0$ oscillating at MHz frequency with velocity $\bm v=v_0 \hat{\bm x} \cos{\omega t}$. The undisturbed (i.e., in the absence of the particle) velocity profiles, $\bm u_0=v_0 \hat{\bm x} \rRe[ \re^{-z/\delta} \re^{-\ri(\omega t-z/\delta)}]$, are shown at two time instants $\omega t\!=\!0$ (solid, red) and $\omega t\!=\!\pi/2$ (dashed, blue) vs. the scaled vertical distance $z/\delta$. The short-dashed vertical line stands for the zero value of the velocity. \label{fig:schematic}}
\end{figure}

In the absence of particles, the horizontal small-amplitude time-periodic oscillations of the resonator at $z\!=\!0$ with velocity $v_0\hat{\bm x} \cos{\omega t}$ create unidirectional oscillatory flow of the viscous liquid of viscosity $\eta$ and density $\rho$ occupying the upper half-space $z\!>\!0$ with velocity given by the real part of $v_0 \hat{\bm x} \re^{-z/\delta} \re^{-\ri(\omega t-z/\delta)}$ \cite{LL}. The flow disturbance propagates upward as the transverse wave attenuated by the exponential factor with $\delta=(2\nu/\omega)^{1/2}$ known as \emph{viscous penetration depth}, where  $\nu\!=\!\eta/\rho$ stands for the kinematic viscosity of the fluid (see Fig.~\ref{fig:schematic}). Computing the shear stress at the resonator, $\sigma_{xz}=\eta\, (\partial u_x/\partial z)_{z=0}$, and dividing by the resonator velocity readily yields the impedance ${\mathcal Z}\!=\!(\ri-1)\, \eta v_0/\delta$ \cite{qcm-liquid}, corresponding to a negative frequency shift and positive dissipation factor (as compared to the unloaded resonator oscillating in vacuum).  Obviously, the particles located above the resonator would perturb this flow and modify the shear stress exerted onto the resonator. The contribution to impedance due to the flow disturbance is entirely \emph{fluid-mediated}, i.e., it takes place for both adsorbed and freely suspended particles, as it does not require a physical contact between the particle and the resonator. For the adsorbed particle, however, there is another contribution to impedance due to the force exerted on its surface by the perturbed flow and transmitted to the resonator via contact. 
 
The prior works applied a variety of numerical methods to account for the hydrodynamics and compute the perturbed viscous stress viscous stress exerted on resonator due to an adsorbed particle. Various factors, such as particle size, surface coverage, particle mobility (e.g., rocking vs. sliding motion), deviation from sphericity and other factors were considered using Finite Element method (FEM) in the early works \cite{johan09,tellech09}, and later with Lattice Boltmann method \cite{lbm1,lbm2,lbm3} and the Immersed Boundary method \cite{ibm20,ibm23}. Although the numerical methods are very powerful, the complex interplay of various factors and uncertainty of physicochemical properties and/or parameters governing the contact dynamics, call for a more analytical approach able to dissect the role of the hydrodynamic forces in QCM-D analysis of discrete adsorbates. In Ref.~\cite{slava18} the hydrodynamic contribution to the impedance due to an adsorbed particle was approximated by the analytical result for the force exerted on a rigid sphere oscillating in an \emph{unbounded} viscous liquid (see, e.g., \cite{LL}). One may expect such approximation to hold for a relatively large (i.e., with respect to the penetration depth $\delta$) particle, as most of its surface is in contact with otherwise quiescent fluid located above the viscous penetration layer. Such assumption, however, requires justification since the unsteady viscous flow in a wall-bounded domain could be quite different from the unbounded flow (e.g., \cite{fl18}). Obviously, for particle of the size comparable to or smaller than the viscous penetration depth, this approximation would not apply. Moreover, the above approximation implicitly assumed that the hydrodynamic contribution to the \emph{contact} force dominates over its fluid-mediated counterpart, which was entirely neglected.   

The fluid-mediated contribution to the QCM-D impedance due to adsorbed particles was recently studied in Ref.~\cite{Buscalioni21} using point-like particle approximation assiming adhesion due to strong lubrication forces. This theory was later revisited in Ref.~\cite{prf23} where the \emph{excess} shear force (or impedance) due to presence of either freely suspended or well adhered (i.e., oscillating as a while with a resonator) discrete particles was determined analytically using a distant-particle asymptotic theory. The derived in Ref.~\cite{prf23} closed-form expressions for the impedance and the velocity (linear and angular) of the freely suspended particle show a very close agreement with the numerical (FEM) computations down to a rather close proximity of less than a particle radius. It was found, in particular, that for some realistic experimental conditions the flow disturbance due to a layer of freely suspended particles located above the resonator, produces the common (``inertial loading") response with $\Delta f<0$ and $\Delta\Gamma>0$ of a magnitude of a few Hz's (at $f_0\!=\!5$~MHz). The same layer of adsorbed particles, however, results in the \emph{positive} frequency shift and unorthodox \emph{negative} bandwidth shift of some hundreds of Hz's. Notice the positive frequency shift (which is typically associated with non-hydrodynamic effects, such as contact viscoelasticity), while $\Delta\Gamma\!<\!0$, implies \emph{reduced dissipation} due to presence of the adsorbed particles. The reason for the seemingly unphysical (sign- and magnitude-wise) response, is that the analysis only concerned the excess shear due to the flow disturbance, whereas an adsorbed particle oscillating with a resonator as a whole excludes a fluid volume above it and also shields the resonator from the transverse shear wave that persists in the particle absence. The \emph{net} excess shear force due to adsorbed particles should, however, combine the fluid-mediated and the contact force. In the present paper we provide a detailed theoretical study of the net excess shear force (impedance) due to discrete adsorbates  at low surface coverage in the analytically tractable limit of a stiff contact, which allows to decouple and analyze the role hydrodynamics independently from other physical phenomena.   \\

\noindent \textit{Problem formulation.} The viscous incompressible liquid in the half-infinite space $z\!>\!0$ is set into motion by the time-periodic horizontal oscillations of the infinite plane at $z\!=\!0$ along the $x$-axis with frequency $\omega$ and amplitude $v_0$ (see Fig.~\ref{fig:schematic}). We further assume that a spherical particle of radius $a$ firmly adheres to the plane and, therefore, oscillates with it in-sync without rotation. Assuming small amplitude of the oscillations, $v_0/\omega \ll a$, to the leading approximation the flow velocity ${\bm V}$ satisfies the unsteady Stokes equations
\begin{eqnarray}&&\!\!\!\!\!\!\!\!\!\!\!\!\!
\partial_t {\bm V}\!=\!-\rho^{-1}\nabla P\!+\!\nu \nabla^2 {\bm V},\ \ \nabla\cdot{\bm V}\!=\!0,
\nonumber\\&&\!\!\!\!\!\!\!\!\!\!\!\!\! {\bm V}(z\!=\!0) = {\bm V}(r\!=\!a, t) = v_0 \hat{\bm x} \cos (\omega t)\,.\label{fud}
\end{eqnarray}
where $P$ is the pressure, $\rho$ and $\nu\!=\!\eta/\rho$ are the density and the kinematic viscosity of the fluid, respectively, and the  spherical distance $r=|\bm x-\bm x_c|$ is measured from the particle center located at $\bm x_c=(0, 0, h)$. Although the particle adhesion corresponds to a vanishing separation distance, $h \! \approx \! a$, we follow the general formulation \cite{Buscalioni21, prf23} and keep an arbitrary proximity $h \geq a$ in the analysis below. We introduce dimensionless variables by normalizing fluid velocity with $v_0$, pressure with $\eta v_0/a$,  time with $\omega^{-1}$ and distance with $a$. Thus the dimensionless (complex) flow field $\bm v$ and pressure $p$ defined via ${\bm V}=v_0\rRe[\re^{-\ri \omega t} \bm v]$ and $P=\eta v_0\rRe[\re^{-\ri \omega t} p]/a$, where $\rRe$ stands for the real part, satisfy
\begin{eqnarray}&&\!\!\!\!\!\!\!\!\!\!\!\!\!
\lambda^2 \bm v\!=\!-\nabla p\!+\!\nabla^2\bm v,\ \ \nabla\cdot\bm v\!=\!0,
\nonumber\\&&\!\!\!\!\!\!\!\!\!\!\!\!\! \bm v(z\!=\!0)= \hat{\bm x},\ \ \bm v(r\!=\!1)=\hat{\bm x}\, . \label{fdon}
\end{eqnarray}
Here $\lambda^2\!=\!-\ri a^2\omega/\nu=-2\ri(a/\delta)^2$. In the absence of a particle, the solution of Eqs.~(\ref{fdon}) is given by $\bm u_0=\re^{-\lambda z}\hat{\bm x}$, where $\lambda\!=\!(1-\ri)\,(a/\delta)$, and $p_0\!=\!0$.

When the particle is present, no analytical solution of Eqs.~(\ref{fdon}) is readily available, however some analytical progress is possible, e.g., for a distant particle (see \cite{prf23}). The major aim of this paper is determining the $x$-component of the complex \emph{excess} shear force (i.e., in excess to the shear force applied by the particle-free background flow), $F$ exerted on the oscillating plate in the incompressible viscous liquid due to an adsorbed particle.

For low values of the particle surface number density, $\tilde n$, the mutual hydrodynamic interactions between particles can be neglected, and the dimensionless excess shear force $F/\eta a v_0$ is equivalent to the dimensionless impedance, ${\!\mathcal Z}/(\eta a \tilde n)$ probed by the QCM-D device. The \emph{net} excess shear force $F$ has two contributions: (i) the fluid-mediated contribution (screening or shielding force) due to presence of the particle and (ii) the direct force the particle exerts on the surface \emph{via contact}. \\

\noindent \textit{Fluid-mediated force.}  The dimensionless stress tensor corresponding to $\{\bm v, p\}$ in Eqs.~(\ref{fdon}) is defined by $\sigma_{ik}\!\equiv\! - p\delta_{ik}\!+\!\partial_k v_i\!+\!\partial_i v_k$. In absence of the particle, $\sigma_{ik}$ has only $xz$ and $zx$ components, which at the plane $z\!=\!0$ equal to $-\lambda$. If the particle is present, it modifies the stress exerted on the resonator by the fluid in the vicinity of the contact, however, far from the particle we shall still have $\sigma_{xz}\! \simeq \!-\lambda$. Therefore, the net fluid-mediated \emph{excess} shear force $F_a$ (i.e., in excess of $-\lambda$ times the surface of the resonator) exerted on the oscillating plate due to presence of an adsorbed particle is defined by
\begin{eqnarray}&&\!\!\!\!\!\!\!\!\!\!\!\!\!
F_a\!=\!\int_{z=0}\! \left(\sigma_{xz}\!+\!\lambda\right) dx dy, \label{ff0}
\end{eqnarray}

The flow perturbation, $\bm u=\bm v-\re^{-\lambda z}\hat{\bm x}$, is governed by:
\begin{eqnarray}&&\!\!\!\!\!\!\!\!\!\!\!\!\!
\lambda^2 \bm u\!=\!-\nabla p\!+\!\nabla^2\bm u,\ \ \nabla\cdot\bm u\!=\!0,
\nonumber\\&&\!\!\!\!\!\!\!\!\!\!\!\!\! \bm u(z\!=\!0)= 0,\ \ \bm u(r\!=\!1)=\left(1-\re^{-\lambda z}\right)\hat{\bm x}. \label{adsorbed}
\end{eqnarray}
The stress tensor $\sigma_{ik}' \!=\!- p\delta_{ik}\!+\!\partial_k u_i\!+\!\partial_i u_k$ corresponding to $\{\bm u, p\}$ in Eq.~(\ref{adsorbed}) obeys $\lambda^2 u_i\!=\!\partial_k \sigma'_{ik}$ and can be written via $\sigma_{ik}$ as
\begin{eqnarray}\!\!\!\!\!\!\!\!\!\!\!\!\!
\sigma_{ik}'\!=\!\sigma_{ik}+\left(\delta_{ix}\delta_{kz}+\delta_{iz}\delta_{kx}\right)\lambda \re^{-\lambda z}\,.\label{st}
\end{eqnarray}
Thus $F_a$ in (\ref{ff0}) can then be written as
\begin{eqnarray}&&\!\!\!\!\!\!\!\!\!\!\!\!\!
F_a\!=\!\int_{z=0} \sigma'_{xz} dxdy=\!\int_{z=0} \partial_z u_{x}dxdy. \label{fofc}
\end{eqnarray}
The direct numerical study of the force using Eq.~(\ref{fofc}) is problematic. The general structure of unsteady Stokes flows generated at the particle surface indicates that, at distances from the boundary greater than the viscous penetration depth $\delta/a\! \propto\! |\lambda|^{-1}$, the flow $\bm u$ a is a superposition of a potential (inviscid) flow and exponential correction, see, e.g. \cite{LL}. However the contribution of the dominant potential flow component into the integral in Eq.~(\ref{fofc}) vanishes identically. Hence $F_a$ is controlled entirely by the exponentially small correction to the potential flow. This renders accurate numerical computation of $F_a$ over infinite plate in Eqs.~(\ref{fofc}) challenging.

We rewrite $F_a$ in the form which is more suitable for the numerical study by using the Lorentz reciprocity \cite{kim}. For an arbitrary incompressible dual flow satisfying $\lambda^2{\hat v}_i=\partial_k{\hat \sigma}_{ik}$  we have:
\begin{eqnarray}&&\!\!\!\!\!\!\!\!\!\!\!\!\!
\frac{\partial ({\hat v}_i \sigma'_{ik})}{\partial x_k}= \frac{\partial (u_i{\hat \sigma}_{ik})}{\partial x_k}. \label{lorentz}
\end{eqnarray}
Integrating Eq.~(\ref{lorentz}) over the fluid volume in the semi-infinite domain, applying the divergence therem and using the original flow field $\bm v$ satisfying Eqs.~(\ref{fdon}) as the dual flow, we find that:
\begin{eqnarray}\!\!\!\!
F_a\!&=&\!-\oint_{r=1}\!\!\re^{-\lambda z}\, \sigma'_{xk} n_k dS\!-\!\frac{4\pi \re^{-\lambda h} (\sinh{\lambda}\!-\!\lambda \cosh{\lambda})}{\lambda}\,
\nonumber\\&&\!\!\!\!
 +\frac{\pi \re^{-2\lambda h} (\sinh{2\lambda}-2\lambda \cosh{2\lambda})}{\lambda}, \label{fa1}
\end{eqnarray}
where we made use of Eq.~(\ref{st}), giving the traction at the particle surface as $\sigma_{xk} n_k\!=\!-\lambda \re^{-\lambda z}\cos{\theta}+\sigma'_{xk} n_k$, where $\theta$ is the polar spherical angle. Thus, instead of integration over the infinite plane at $z\!=\!0$ in Eq.~(\ref{fofc}), the excess shear force $F_a$ can be alternatively evaluated by integrating the traction $\sigma'_{xk} n_k$ over the particle surface at $r\!=\!1$. Notice also that the last two (analytical) terms in the RHS of Eq.~(\ref{fa1}) comprise (up to a factor of $\pi$) the net hydrodynamic contribution to the impedance due to an adsorbed particle reported in Ref.~\cite{Buscalioni21}. The numerical results indicate that the 1st (integral) term is usually dominant over the last two (analytical) terms. \\

\noindent \textit{Contact force and torque.} For \emph{freely suspended} particles the excess shear force exerted on the resonator is mediated solely by the suspending fluid \cite{prf23}. The adsorbed particle not only modifies the flow above the resonator (i.e., via $F_a$), but also applies a force via \emph{contact}. We assume that the contact force the rigidly attached particle exerts on the plane, $F_c$, is equal in magnitude and opposite in sign to the force that the plane exerts on the particle (see also \cite{ibm23}). The contact force $F_c$ is determined from the Newton's force balance:
\begin{eqnarray}&&\!\!\!\!\!\!\!\!\!\!\!\!\!
\lambda^2 \xi U\!=\! \oint_{r=1}\!\! \sigma_{xk} n_k dS-F_c, \label{eq:fc0}
\end{eqnarray}
where for a particle moving with a plane as a whole its dimensionless translation velocity $U\!=\!1$ and the traction $\sigma_{ik} n_k$ corresponds to the original flow in Eqs.~(\ref{fdon}). Here the parameter $\xi=m/\rho a^3$, where $m$ stands for the particle's mass, characterizes the solid inertia.

Substituting the traction at the particle surface $\sigma_{xk} n_k\!=\!-\lambda \re^{-\lambda z}\cos{\theta}+\sigma'_{xk} n_k$ into Eq.~(\ref{eq:fc0}) yields the following result:
\begin{eqnarray}
F_c &=& -\frac{4\pi\, \re^{-\lambda h} (\sinh{\lambda}-\lambda \cosh{\lambda})}{\lambda}+ \nonumber \\
&& +\oint_{r=1}\!\! \sigma'_{xk} n_k dS -\lambda^2 \xi = F_c'-\lambda^2 \xi\,, \label{fc}
\end{eqnarray}
where $F'_c$ is the hydrodynamic part of the contact force. The net excess shear force due to an adsorbed particle can now be found as $F=F_a+F_c$. Notice that upon neglecting the hydrodynamics entirely, the net excess force is due to inertial mass of the particle, $F\!=\!-\lambda^2\xi\!=\!-(4\pi/3)(\rho_s/\rho)\lambda^2\!=\!\ri m\omega/\eta a$, being equivalent to the (dimensional) impedance ${\mathcal Z}=\ri\omega {\tilde n} m$. Substituting $\omega=2\pi n f_0$ (where $n$ is the overtone number) and using the small-load approximation for the above impedance, we readily arrive at the classical Sauerbrey equation \cite{review2015}:
\begin{equation}
\frac{\Delta f}{f_0} \simeq \ri \frac{\mathcal Z}{\pi {\mathcal Z}_q}=-n\, \frac{{\tilde n} m}{m_q}\;,\label{sauerbrey}
\end{equation}
where $m_q\!=\!\mathcal{Z}_q/(2 f_0)$ and $m {\tilde n}$ is the areal mass density (both have units of mass per unit area).

The contact \emph{torque} $L_c$ (the $y$-component, scaled with $\eta a^2 v_0$) the adsorbed particle exerts on the resonator could also be of interest towards estimating the stiffness of the contact and it is given by (with respect to the particle center at $z\!=\!h$): 
\begin{eqnarray}
\frac{2}{5} \lambda^2  \xi \Omega &=& \oint_{r=1}\!\! \left[(z-h)\sigma_{xk}\!-\!x\sigma_{zk}\right]n_k dS - L_c\,,\label{torque}
\end{eqnarray}
where $\Omega$ is the dimensionless angular velocity of the particle scaled with $v_0/a$. For an adsorbed particle with a stiff contact (i.e., without rotation, $\Omega\!=\!0$) there is no contribution of the solid inertia in the LHS of Eq.~(\ref{torque}) and the contact torque reduces to
\begin{eqnarray}
L_c=\oint_{r=1}\!\! \left[(z-h)\sigma_{xk}\!-\!x\sigma_{zk}\right]n_k dS\,.\label{Lc0}
\end{eqnarray}
The contact torque in Eq.~(\ref{Lc0}) be rewritten as an integral over the perturbed traction $\sigma'_{ik} n_k$ using Eq.~\ref{st} as (cf. Eq.~(22) for $\mathcal B$ in \cite{prf23}):
\begin{eqnarray}
L_c &=& \!\oint _{r=1}\!\!\left[\cos{\theta}\, \sigma'_{xk} \!-\!\sin{\theta} \cos{\phi}\, \sigma'_{zk} \right] n_k dS
\nonumber\\
&&\! -4\pi \re^{-\lambda h}\left[\sinh{\lambda} +\frac{3\left(\sinh{\lambda}-\lambda\cosh{\lambda}\right)}{\lambda^2}\right].
\label{Lc1}
\end{eqnarray}
 If contact torque with respect to the point of contact (at $z\!=\!0$) is considered, then we readily have:
\begin{eqnarray}
L^\mathrm{(c)}_c &=&\!\oint_{r=1}\!\!\left(z\sigma_{xk}\!-\!x\sigma_{zk}\right) n_k dS = \nonumber\\
&& L_c+h\oint_{r=1}\!\!\sigma_{xk} n_k dS = L_c+h F'_c\,,  \label{Lc2}
\end{eqnarray}
where $F_c'$ is the hydrodynamic part of the contact force in Eq.~(\ref{fc}).

Notice that the above derivations of $F_a$ and $F_c$ are rigorous and do not involve any approximation, besides from the assumption of small-amplitude oscillations that allowed to neglect the nonlinear inertia terms in the flow equations. The resulting expressions involve integrals of the traction associated with the perturbed flow, $\sigma'_{ik} n_k$, over the particle surface at $r\!=\!1$ which can be performed numerically. \\

\noindent \textit{Small-particle limit.}  Let us consider the small-particle (or low-frequency) limit, $|\lambda|\ll 1$, for which the steady Stokes equations hold to the first approximation, as the unsteady term $\lambda^2\bm u$ in Eqs.~(\ref{adsorbed}) produces $o(|\lambda^2|)$ corrections in the solution \cite{note0}. We next expand the perturbed flow $\bm u$ in Eqs.~(\ref{adsorbed}) as $\bm u=\lambda \bm u_1+\lambda^2 \bm u_2 + \ldots$.  
At the leading order we have:
\begin{eqnarray}&&\!\!\!\!\!\!\!\!\!\!\!\!\!
-\nabla p_1\!+\!\nabla^2\bm u_1\!=\!0,\ \ \nabla\cdot\bm u_1\!=\!0,
\nonumber\\&&\!\!\!\!\!\!\!\!\!\!\!\!\! \bm u_1(z\!=\!0)= 0,\ \ \bm u_1(r\!=\!1)=z\hat{\bm x}, \label{adsorbedl}
\end{eqnarray}
where $p_1$ stands for pressure to order $\lambda$. Notice that the analytical terms in the r.h.s. of Eqs.~(\ref{fa1}) and (\ref{fc}) are all $\mathcal{O}(|\lambda|^2)$, meaning that at the leading order $\mathcal{O}(|\lambda|)$ we have:
\begin{eqnarray}&&\!\!\!\!\!\!\!\!\!\!\!\!\!
F_c^{(1)}\!=\!-F_a^{(1)}\!=\!\oint_{r=1}\sigma^{\prime (1)}_{xk} n_k dS\,.\label{oneill}
\end{eqnarray}
In other words, \textit{for small particles, the fluid-mediated contribution is compensated by the (hydrodynamic part of) contact force at the leading approximation in $\lambda$}, such that the net excess force due to an adsorbed particle $F\!=\!F_a+F_c$ reduces to $\mathcal{O}(|\lambda|^2)$. The Eqs.~(\ref{adsorbedl}) govern the problem of a steady linear shear flow past a fixed sphere in contact with a plane wall, and its exact solution using special ``touching sphere" coordinates is given in \cite{oneill68}. In particular, the dimensionless contact force to the leading approximation found from (\ref{oneill}) is given by $F_c\!=\!-F_a\!=\!\!-6\pi f \lambda+\mathcal{O}{(|\lambda|^2)}$, where the constant $f\!\simeq\! 1.701$.

Analogously, at $|\lambda|\ll 1$,  the torque applied on the adsorbed particle can be estimated: the 2nd (analytical) term in (\ref{Lc1}) is of ${\mathcal O}(|\lambda|^3)$ and the integral term to the leading approximation contributes $L_c\! \approx\! -4\pi g \lambda$ \cite{note1}, where the constant $g\!\simeq \!0.944$ \cite{oneill68}. Given the asymptotic behavior of $F_c$ above we readily find that at contact ($h\!=\!1$) the torque with respect to the point of contact to the leading approximation redas $L^{(\mathrm{c})}_c \approx -(6f+4g) \pi \lambda\!=\!-13.981 \pi \lambda$.

At the sub-leading order at $\mathcal{O}(|\lambda|^2)$ we have the following problem:
\begin{eqnarray}&&\!\!\!\!\!\!\!\!\!\!\!\!\!
-\nabla p_2\!+\!\nabla^2\bm u_2\!=\!0,\ \ \nabla\cdot\bm u_2\!=\!0,
\nonumber\\&&\!\!\!\!\!\!\!\!\!\!\!\!\! \bm u_2(z\!=\!0)= 0,\ \ \bm u_2(r\!=\!1)=- z^2\hat{\bm x}/2, \label{oneill2}
\end{eqnarray}
The solution of Eqs.~(\ref{oneill2}) that would allow determining the subleading corrections to $F_{a}^{(2)}$ and $F_{c}^{(2)}$ is possible following the analysis in \cite{oneill68}.  

\begin{figure*}[tbh!]
\begin{tabular}{cc}
\includegraphics[width=0.42\textwidth]{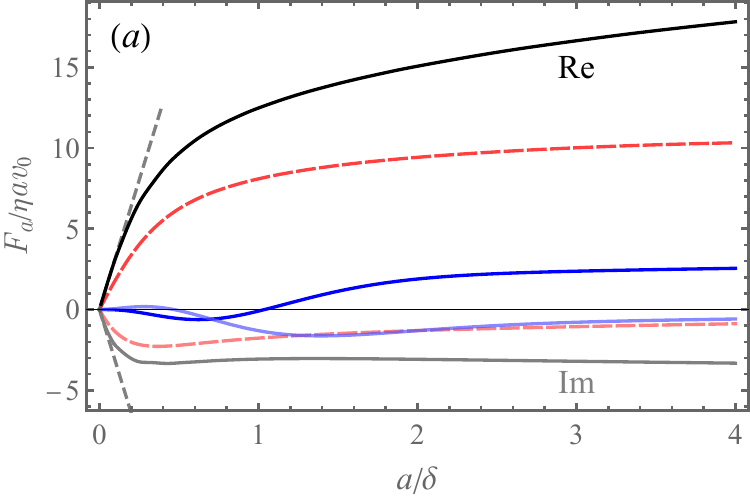} \hspace{3mm} & 
\includegraphics[width=0.44\textwidth]{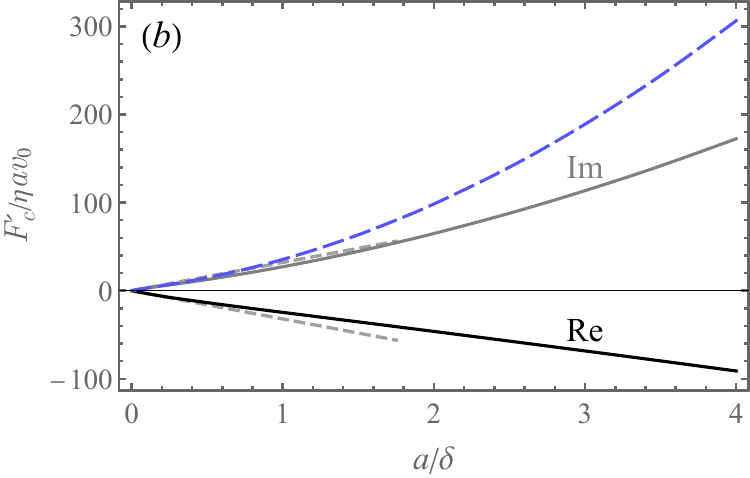} \\ \\
\includegraphics[width=0.44\textwidth]{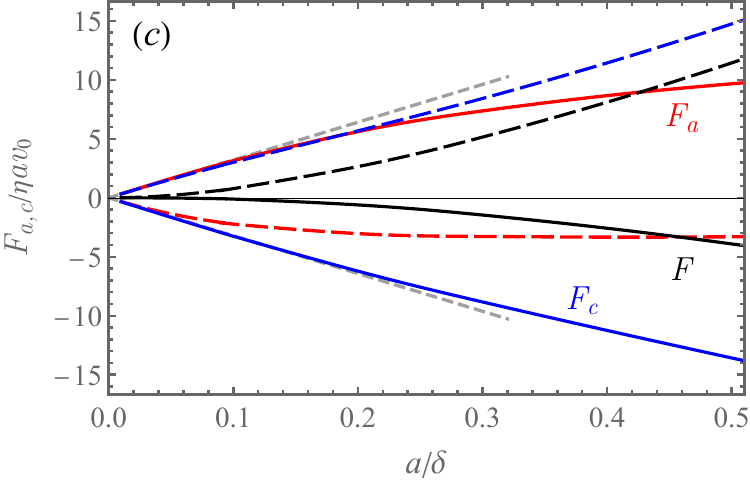}  \hspace{3mm}
 & 
\includegraphics[width=0.44\textwidth]{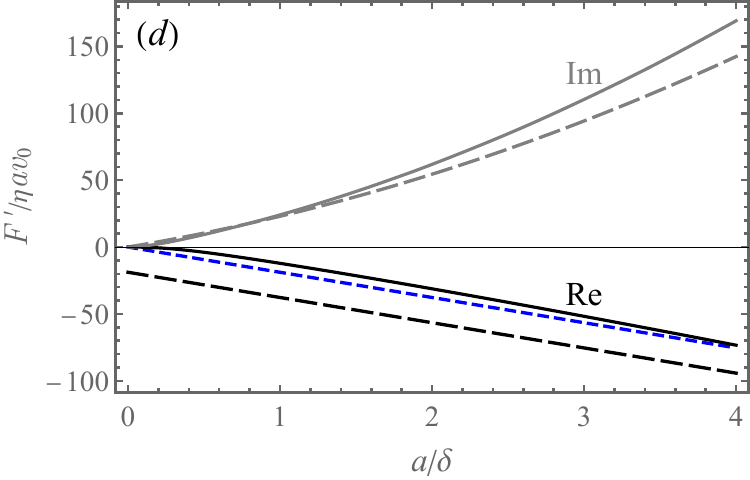}
\end{tabular}
\caption{Excess shear force (real and imaginary part) due to adsorbed particle ($h\!=\!a$) vs. $a/\delta$. a) Fluid-mediated contribution $F_a$: the solid (black) lines stand for the numerical results, short-dashed (gray) lines for the small-$\lambda$ asymptote, $F^{(1)}_a$ and long-dashed (red
) lines correspond to the distant-particle prediction $F_a^\mathrm{asym}$ at $h\!=\!a$ in Eq.~(\ref{Fadist}); the blue curves stand to the analytical part (last two terms) of $F_a$ in Eq.~(\ref{fa1}); b) Hydrodynamic part of the contact force $F'_c$ (black, gray); the short dashed lines for the small-$\lambda$ asymptote $F^{(1)}_c$ and long-dashed (blue) line for imaginary part of the net contact force, $\rIm[F_c]$, (the real part unchanged) for neutrally buoyant particle with $\xi=4\pi/3$; c) Various components of the excess force for $a/\delta\! \lesssim \! 0.5$: $F_c$ (blue, for $\xi=4\pi/3$), $F_a$ (red) and $F$ (black); solid and long-dashed lines stand for real and imaginary parts of different terms, respectively. d) Comparison of the net excess force $F'$ (excluding solid inertia, solid lines) vs. the analytical result $F_0$ \cite{LL} for a sphere oscillating in an unbounded liquid (long-dashed lines); short-dashed (blue) curve stands for the real part of  $F_0$ upon  subtracting the pseudo-Stokes drag term ($-6\pi$) in Eq.~(\ref{eq:LL}). \label{fig:results}}
\end{figure*}

\begin{figure}[tbh]
\includegraphics[width=0.9\columnwidth]{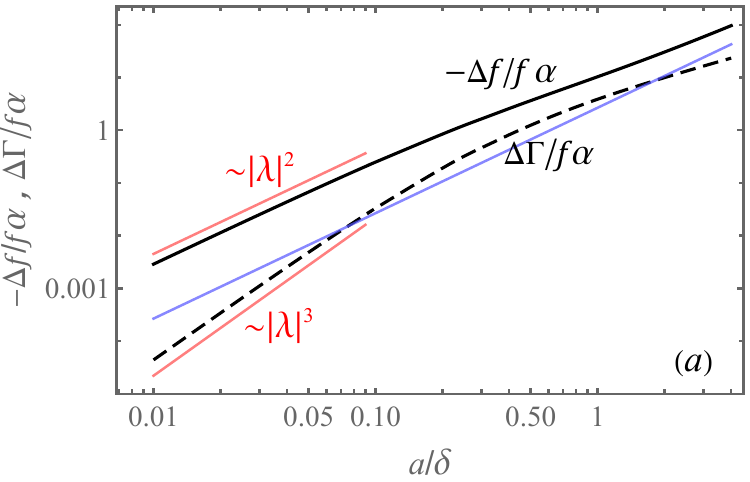} \\
\vspace{2mm}
\hspace{2mm} \includegraphics[width=0.85\columnwidth]{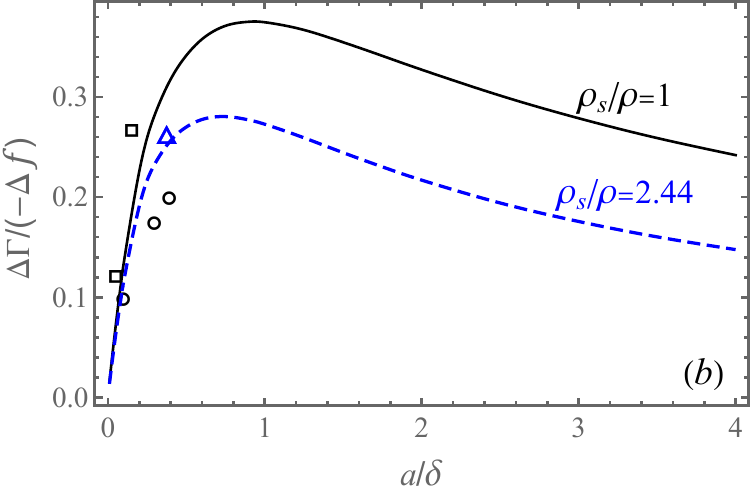}
\caption{a) The dimensionless frequency shift $-\Delta f/(f_0\alpha)$ (solid black curve) and the half-bandwidth $\Delta \Gamma/(f_0\alpha)$ (dashed curve) shift due to adsorbed neutrally buoyant ($\rho_s/\rho\!=\! 1$) particles vs. $a/\delta$ (double-log plot); the red lines designate the asymptotic behavior at $a/\delta\ll 1$; the solid blue line is the Sauerbrey frequency shift, $-\Delta f_S/(f_0 \alpha)$; b) The dimensionless acoustic ratio $\Delta \Gamma/(-\Delta f)$ vs. $a/\delta$ for neutrally buoyant (solid line) and non-buoyant (dashed line) particles with $\rho_s/\rho\!=\! 2.44$ (e..g, silica nanoparticles in ethanol \cite{grun15}); empty squares ($\square$) are the results for $26$~nm and $73$~nm diameter polystyrene nanoparticles at fundamental frequency\cite{adam20}, circles ($\circ$) are the results for 30~nm CPMV particles and $86$~nm and $114$~nm liposomes (at the 3rd overtone)\cite{tellech09}  and $\vartriangle$ stand for the 137~nm silica nanoparticles adsorbing on gold from ethanol (at the 3rd overtone)\cite{grun15}.  \label{fig:shifts}}
\end{figure}

Thus, finding the net excess force $F$ at the lowest non-trivial order demands the solution of the following problem at  $\mathcal{O}(|\lambda|^2)$:
\begin{eqnarray}&&\!\!\!\!\!\!\!\!\!\!\!\!\!
-\nabla p_2\!+\!\nabla^2\bm u_2\!=\!0,\ \ \nabla\cdot\bm u_2\!=\!0,
\nonumber\\&&\!\!\!\!\!\!\!\!\!\!\!\!\! \bm u_2(z\!=\!0)= 0,\ \ \bm u_2(r\!=\!1)=- z^2\hat{\bm x}/2, \label{oneill2}
\end{eqnarray}
The analytical solution of Eqs.~(\ref{oneill2}), that would allow determining the subleading corrections to $F_{a}^{(2)}$ and $F_{c}^{(2)}$, is possible following the analysis in \cite{oneill68}. However, using Eqs.~(\ref{fa1}) and (\ref{fc}) it can be shown, that the net excess shear force to the leading approximation is yet determined by the stress perturbation in Eqs.~(\ref{adsorbedl}):
\begin{equation}
F=F_a+F_c\! \approx \! \lambda^2 \oint_{r=1} z \sigma^{(1)\prime}_{xk} n_k dS-\lambda^2 \xi\,. \label{netforce}
\end{equation}
Notice that since $\lambda^2\!=\!-\ri a^2\omega /\nu$ is purely imaginary, the leading contribution to the real part of $F$ reduces to ${\mathcal O}(|\lambda|^3)$, implying that for small particles $\Delta\Gamma$ is expected to be smaller than $\Delta f$ (see Figs.~\ref{fig:results}c and \ref{fig:shifts}a below). Computing the net excess shear force from Eq.~(\ref{netforce}) shall be performed  elsewhere.  
\\

\noindent \textit{Numerical computations.}
The numerical solution of Eqs.~(\ref{adsorbed}) is performed in the dimensionless cylindrical coordinates $\{\varrho,\phi,z\}$ (all distances scaled with $a$),
such that $x\!=\!\varrho \cos\phi$, $y\!=\!\varrho \sin\phi$, with its origin at the plate at $z=0$ and the $z$-axis passing through the center of the adsorbed spherical particle.
\begin{figure}[tbh]
\includegraphics[width=0.85\columnwidth]{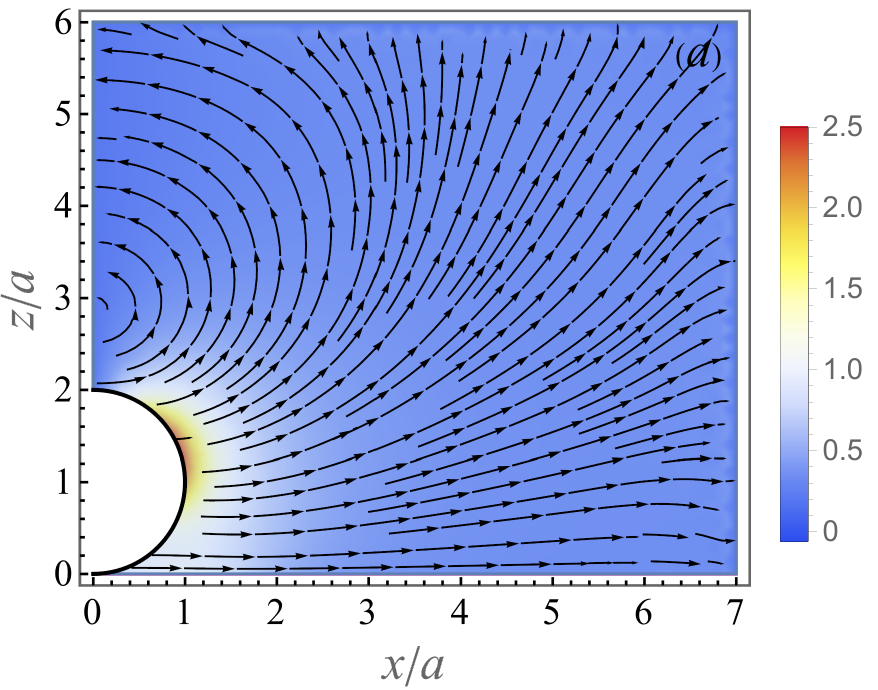}  \\
\vspace{2mm}
\includegraphics[width=0.85\columnwidth]{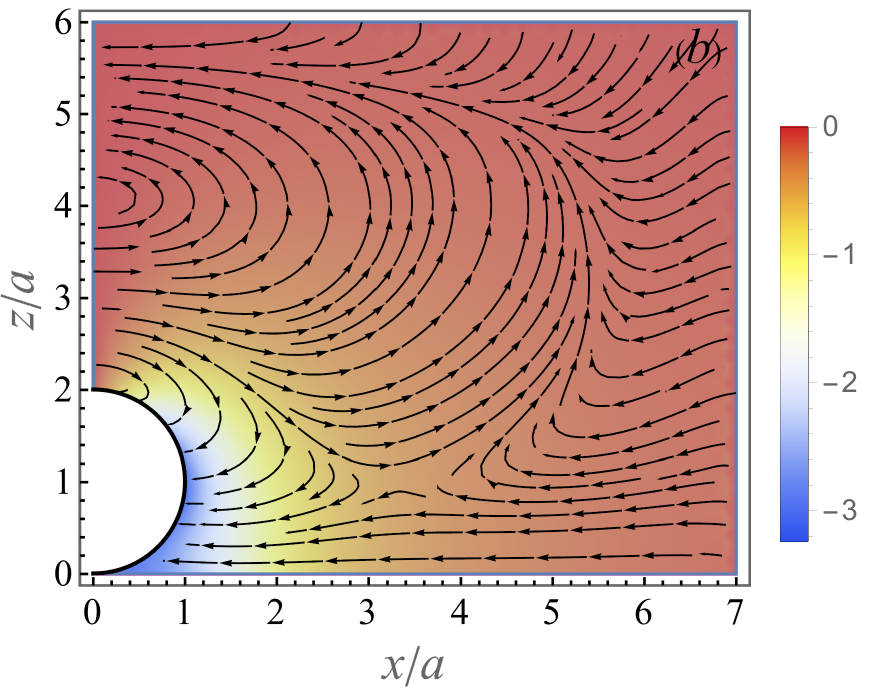}
\caption{The perturbed flow (streamlines) and pressure (color map) fields in Eq.~(\ref{fdon}) due to an adsorbed particle for $\delta/a\!=\!1$ in $xz$-plane (for $\phi\!=\!0$) at two different time instances: (a) velocity $\{\rRe [\mathcal U], \rRe [\mathcal W]\}$ and pressure $\rRe[\mathcal P]$ corresponding to $\omega t=0$; (b) velocity $\{\rIm [\mathcal U], \rIm [\mathcal W]\}$ and $\rIm[\mathcal P]$ corresponding to $\omega t=\pi/2$. \label{fig:flow}}
\end{figure}
We use the following ansatz admitting simple dependence on the azimuthal angle:  $v_{\varrho}\!=\!{\mathcal U}(\varrho, z)\cos{\phi}$, $v_{\phi}\!=\!{\mathcal V}(\varrho, z)\sin{\phi}$, $v_z\!=\! {\mathcal W}(\varrho, z)\cos{\phi}$ and $p={\mathcal P}(\varrho, z)\cos{\phi}$ which reduces the solution to two dimensions \cite{prl20, prf23}. The corresponding problem for $\mathcal U$, $\mathcal V$, $\mathcal W$ and $\mathcal P$ is defined in the rectangular domain $0\le\varrho\le\varrho_{m},\; 0\le z \le z_m$ with an exclusion of the half unit disk centered at $(0, h)$ representing the adsorbed particle. The pressure $\mathcal P$ is set to a fixed (zero) value far from the particle at $z\!=\!z_\mathrm{max},\ \varrho\!=\!\varrho_\mathrm{max}$. The boundary condition $\bm u\!=\!0$ is applied at $\varrho\!=\!\varrho_\mathrm{max}$,  $z\!=\!0$ and $z\!=\!z_\mathrm{max}$. We set no-flux boundary condition at $\varrho\! =\!0$, while at the boundary of half-circle we specify $\mathcal U\! =\! -\mathcal V\!=\! 1-\re^{-\lambda z}$ and $\mathcal W \!=\! 0$. We then apply the Finite Element Method (FEM) implemented in {\it Mathematica} 12.0 to solve the Eqs.~(\ref{adsorbed}).  A typical mesh size is selected to be $0.05$ within the domain and $0.025$ along the boundaries. Notice that for stiff contact, the particle is oscillating in-sync with the resonator and there is no relative shearing (or sliding) motion between the two. In Ref.~\cite{prf23} the fluid-mediated part of the excess shear force ($F_a$) for an adsorbed particle was determined via the numerical solution of the auxiliary problem corresponding to a stationary (heavy inertial) particle located above the resonator, and this resulted in numerical difficulties at close proximity owing to strong lubrication forces. The direct formulation of the problem in Eqs.~\ref{adsorbed} circumvents these complications, allowing for accurate numerical solution near contact, $h \! \rightarrow \! 1$.

Numerical computation shows that the flow $\bm u$ converges at $\varrho_\mathrm{max}\!\simeq\!9$, $z_\mathrm{max}\!\simeq\!9+h$. The typical flow and pressure disturbance due to an adsorbed particle for $\delta\!=\!1$ and $h\!=\!1.001$ in meridional plane $xz$--plane (for $\phi\!=\!0$) are shown in Figs.~\ref{fig:flow}a,b at two instances, $\omega t\!=\!0$ and $\omega t\!=\!\pi/2$, respectively. It can be readily seen, that the interaction of the transverse wave originated at the oscillating plate (see the undisturbed velocity in Fig.~\ref{fig:schematic}) with the particle creates a rather complex flow pattern with transient recirculations. \\

\noindent \textit{Results and discussion}
The numerical results for the real and imaginary part of the excess shear force due to an adsorbed particle at contact ($h\!=\!a$) are presented in Figs.~\ref{fig:results}a-d (solid curves).
The fluid-mediated contribution $F_a$ in Eq.~(\ref{fa1}) is depicted in Fig.~\ref{fig:results}a vs. $a/\delta$ together with the linear small-$\lambda$ asymptotes $F^{(1)}_a$ (short-dashed lines) and the prediction of the the distant-particle theory (long-dashed, red curves) that assumes $h\!\gg\! \mathrm{max}(a,\delta)$, while the ratio $a/\delta$ is not constrained \cite{prf23}:
\begin{eqnarray}
F_a^\mathrm{asym} &=&6\pi\re^{\lambda (1-h)} -\frac{\pi^2 \re^{-2\lambda h} }{\lambda}\times \nonumber\\
&&  \left[\frac{3(\re^{2\lambda}-1)}{\pi}+\sum_{l=1}^{\infty}\frac{4(l\!+\!1)I_{l+1/2}(\lambda)}{K_{l+1/2}(\lambda) }\right]\,. \label{Fadist}
\end{eqnarray}
Here $I_{\nu}(\lambda)$ and $K_{\nu}(\lambda)$ are the modified Bessel functions of the 1st and 2nd kind, respectively. It can be readily seen that the numerical results show an excellent agreement with $F^{(1)}_a$ at low values of $a/\delta$. The agreement with the theoretical prediction in Eq.~(\ref{Fadist}) is only qualitative. Recall that starting from relatively small separations, $h\gtrsim 1.5a$, a surprisingly close agreement between the numerical results and Eq.~(\ref{Fadist}) was found \cite{prf23}, while at contact ($h\!=\!a$) the theory considerably underestimates the fluid-mediated contribution, $F_a$ (i.e., both the real and the imaginary parts, see red long-dashed curves in Fig.~\ref{fig:results}a). Another observation is that the relative weight of the analytical (the last two) terms in Eq.~(\ref{fa1}) to $F_a$ is small for all values of $a/\delta$ (see the blue curves in Fig.~\ref{fig:results}a).

Notice that $\rRe[F_a]>0$ while $\rIm[F_a]<0$, which implies positive frequency shift (which is typically associated with non-hydrodynamic effects, such as contact viscoelasticity), and $\Delta\Gamma\!<\!0$, indicating \emph{reduced dissipation}. The reason for seemingly unorthodox result, is that the adsorbed particle excludes a fluid volume above the resonator and in the same time shields the resonator from the shear wave that would otherwise persist in its absence. One might expect, that adding the contact force would flip the signs of the net excess force (see below).

The numerical results for the hydrodynamic part of the contact force (excluding solid inertia), $F'_c$ [the sum of the first two terms in Eq.~(\ref{fc})], are depicted in Fig.~\ref{fig:results}b vs. $a/\delta$ (solid curves). The linear small-$\lambda$ asymptotes $F^{(1)}_c$ (short-dashed lines) approximate $F_c'$ very well up to $a/\delta \approx 1$. The long-dashed (blue) line stands for the net contact force $F_c$ in Eq.~(\ref{fc}) for neutrally buoyant particle with $\xi=4\pi/3$. It can be readily seen, that for $a\gtrsim \delta$ the excess force is dominated by the contact force, as $F_c \gg F_a$, while for $a/\delta \lesssim 0.5$, the two terms are comparable. Moreover, since $F^{(1)}_a=-F^{(1)}_c$, their contributions compensate each other and the net effect is $\mathcal{O}(|\lambda|^2)$. This notion is illustrated in Fig.~\ref{fig:results}c, where we plot $F_a$, $F_c$ (for neutrally buoyant particle, $\xi=4\pi/3$) and the resulting net excess force $F$ vs. $a/\delta<0.5$. The small-$\lambda$ linear asymptotes are shown as short dashed lines. The exact cancelation of the fluid-mediated and contact forces at the leading order in $\lambda$ result in rather low values of $F$ for small particle, in particular its real part of ${\mathcal O}(|\lambda|^3)$, while the imaginary part is of ${\mathcal O}(|\lambda|^2)$ (see the analysis above). For example, for 50~nm ($a/\delta\!=\!0.1$) neutrally buoyant particles in water for the fundamental frequency of $f_0\!=\!\omega/2\pi\!=\!5$~MHz, giving $\delta \approx 252$~nm), yields ${\mathcal Z}/(\eta a \tilde n)\!\approx\! -0.10+0.78 \ri$. Using the small-load approximation \cite{johan21}, the shift in oscillation frequency, $\Delta f$, and in its half-bandwidth, $\Delta \Gamma$ (related to the dissipation factor, $\Delta {\mathcal D}\!=\!2\Delta\Gamma/f$), can be found from $\Delta f-\ri \Delta\Gamma\!\simeq \!\ri f_0 {\mathcal Z}/(\pi \mathcal{Z}_q)$, where the quartz resonator's shear wave impedance $\mathcal{Z}_q\!=\!8.8\cdot 10^6$~$\mathrm{kg}/\mathrm{m}^2 \mathrm{s}$ and the oscillation frequency $f\!=\!n f_0$ where $n\!=\!1,3,5,\dots$ is the overtone number. Assuming the particle number density at the surface of the resonator $\tilde n\!=\!0.01 a^{-2}$ (i.e., one nanoparticle per $100 a^2$ surface area), the small-load approximation at the fundamental frequency $f_0$ yields $\Delta f\!\approx \!-56.0$~Hz and $\Delta\Gamma\!\approx \!7.4$~Hz only.

In Fig.~\ref{fig:results}d we compare the hydrodynamic part of the net excess shear force, $F'$ (excluding the solid inertia term, solid curves) vs. the classical result for the force exerted on an rigid sphere oscillating with velocity $\bm u_0= v_0 \hat{\bm x}\re^{-\ri\omega t}$ in an \emph{unbounded} viscous liquid, quiescent at infinity (long-dashed lines).  This force can be written in the dimensionless form (scaled with $\eta a v_0$) as  (see, e.g., \cite{LL}):
\be
F_{0}=-6\pi\left(1+\frac{a}{\delta}\right)+6\pi\ri \left(\frac{a}{\delta}\right) \left(1+\frac{2}{9}\frac{a}{\delta} \right)\,. \label{eq:LL}
\ee
It was previously proposed \cite{slava18}, that for large enough particles ($a\! \gg \!\delta$), the hydrodynamic contribution to the impedance can be closely approximated by $F_{0}$, as most of the particle surface oscillates in almost quiescent liquid located above the penetration depth $\delta$. It can be seen that the agreement between numerical result for $\rIm[F']$ (dashed line) and the 2nd (``added mass") term in Eq.~(\ref{eq:LL}) is quite close and the relative error (which increases with $a/\delta$) is $\sim\!16$~\% for $a/\delta\!=\!4$. For the same value of $a/\delta$, the real part, $\rRe[F']$ deviates from the 1st (``drag") term in Eq.~(\ref{eq:LL}) by $\sim\! 22$\%, while this error becomes larger for smaller particles, e.g., it is already $\sim\!68$\% for $a/\delta\!=\!1$. It appears that subtracting the zero-frequency pseudo-Stokes drag term $-6\pi$ from $\rRe[F_0]$ yields much closer agreement (see the short-dashed line in Fig.~\ref{fig:results}d), in particular for large values of $a/\delta$. For instance, for $a/\delta\!=\!4$ the error is only $\sim\!2.7$\%, while for $a/\delta\!=\!1$ the error is $\sim\!35$\%. \\

The dimensionless frequency shift, $-\Delta f/(f_0 \alpha)$ and half bandwidth shift, $\Delta \Gamma/(f_0 \alpha)$ vs. $a/\delta$ for neutrally buoyant particles (i.e., $\rho_s/\rho\!=\! 1$) particles are shown as log-log plot in Fig.~\ref{fig:shifts}a (see the black solid and dashed curves). Here $\alpha\!=\! \eta a {\tilde n}/\mathcal{Z}_q$ is the dimensionless (viscous-to-solid) impedance ratio. For example, for $50$~nm-diameter particles in water and particle surface density $\tilde n\!=\!0.01 a^{-2}$, we find that $\alpha\!=\!4.55\cdot 10^{-5}$. Both shifts are monotonically increasing functions of $a/\delta$, while for small values of $a/\delta$ we have $-\Delta f \! \propto\! (a/\delta)^2$ and $\Delta \Gamma \! \propto \! (a/\delta)^3$. The scaled frequency shift due to \red{the particle inertia alone by the Sauerbrey Eq.~(\ref{sauerbrey})}, $-\Delta f_S/(f_0 \alpha)\! = \!\frac{8}{3} (\rho_s/\rho) (a/\delta)^2$ for neutrally buoyant particles is depicted for comparison (solid gray line). It can be readily seen that the Sauerbrey equation significantly \emph{underestimates} the mass of finite size adsorbates . The dimensionless acoustic ratio, $\Delta \Gamma/(-\Delta f)$ is independent of the surface coverage $\tilde n$ (provided it is low enough so that hydrodynamic interaction between distinct adsorbed particles can be neglected) and oscillation frequency; it is shown vs. $a/\delta$ in Fig.~\ref{fig:shifts}b (solid line) together with some published experimental results (symbols). Notice that for adsorbed particles with stiff contact the theory predicts that the acoustic ratio is bounded, e.g., for neutrally buoyant particles $\Delta \Gamma/(-\Delta f) \lesssim 0.38$. Heavier particles are expected to yield even smaller values of the acoustic ratio at the peak, as the inertial (Sauerbrey) term $-\lambda^2 \xi$ in (\ref{fc}) contributes to the imaginary part of $F$ therefore increases the (negative) frequency shift, $(-\Delta f)$, while $\Delta\Gamma$ remains unchanged. For instance, for silica nanoparticles  $\rho_s\!=\!1.93$~g/cm$^3$ suspended in ethanol ($\rho\!=\!0.79$~g/cm$^3$) \cite{grun15} we have $\Delta \Gamma/(-\Delta f) \lesssim 0.28$.

\begin{figure}[t]
\includegraphics[width=0.85\columnwidth]{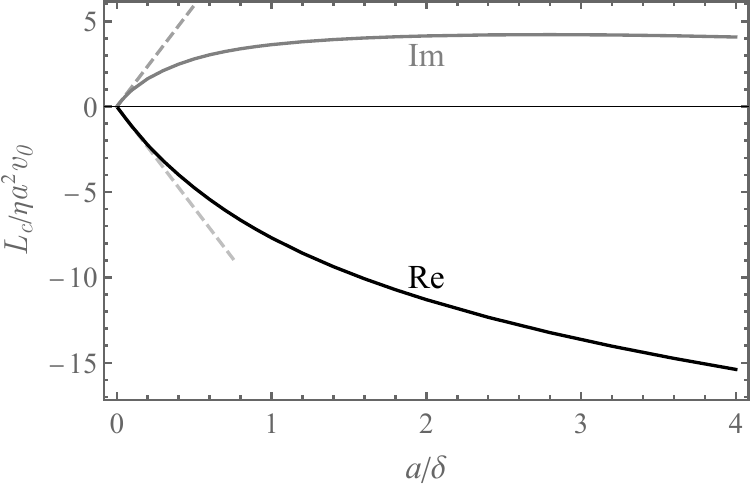}
\caption{The contact torque $L_c/\eta a^2 v_0$ with respect to the particle center vs. $a/\delta$. Solid (black) and long-dashed (gray) curves stand for the numerical results for the real and the imaginary parts of $L_c$;  the short-dashed (gray) lines stands for the small-$\lambda$ asymptotics. \label{fig:torque}}
\end{figure}

Finally, in Fig.~\ref{fig:torque} the real and imaginary parts of the contact torque $L_c/\eta a^2 v_0$ (with respect to the particle center at $z\!=\!h$ in Eq.~\ref{Lc1}) is plotted vs. $a/\delta$. The small-$\lambda$ asymptotic (short-dashed lines) show an excellent agreement with the numerical results (black solid and gray long-dashed curves). 
Notice that the torque $L_c^{(c)}$ with respect to the \emph{point of contact} in Eq.~(\ref{Lc2}) would be much higher owing to the large contact force, since 
$|a F_c'|\! \gg\! |L_c|$. \\

\noindent \textit{Conclusions and perspectives.}
Fluid-mediated contribution to the excess shear force and the hydrodynamic part of the contact force are in competition for $a/\delta \lesssim 0.5$, while the net effect of the viscous stresses reduces to ${\mathcal O}(|\lambda|^2)$ (and to ${\mathcal O}(|\lambda|^3)$ for the dissipation factor) due to the mutual cancellation of the linear in $\lambda$ terms in $F_a$ and $F_c'$. Since the Sauerbrey contribution due to particle solid inertia is also of $O(|\lambda|^2)$, it implies that accurate account of hydrodynamics in the analysis of QCM-D response is equally important. \\

We have previously shown that in the limit of vanishing proximity, $\epsilon\!=\!h/a-1 \! \to \! 0$, the translation and rotation velocities of a freely suspended spherical particle to the leading approximation in $\epsilon$ tend to that of the rigidly attached particle, i.e., $V-1, \Omega \sim |\ln{\epsilon}|^{-1}$ solely due to the lubrication forces (see Sec.~V and Eq.~90 in \cite{prf23}). However, despite this fact, the fluid-mediated contribution to the excess shear force due to freely suspended particle, that can be written as $F_f\!=\!F_a+(V-1){\mathcal A}+\Omega {\mathcal B}$, is different from the corresponding contribution due to an adsorbed particle, $F_f\!\ne\! F_a$, due to divergence of the corresponding resistance functions, $\mathcal A$ and $\mathcal B$ at $\epsilon\! \to \! 0$. 
This argument explains the apparent disagreement between the prediction of the impedance due to the freely suspended particles \emph{near contact} and the adsorbed particle \emph{at contact}. Similarly, arguments concern the contact force $F_c$ in Eq.~(\ref{eq:fc0}), which is zero for a freely suspended particle and takes a finite value (i.e., due to the hydrodynamics) for a particle forming a stiff contact with the resonator.

In the limit of low surface coverage and stiff contact both shifts, $-\Delta f/(\alpha f_0)$ and $\Delta\Gamma/(\alpha f_0)$, are expected to increase with  the oscillation frequency (i.e., the overtone number $n$) as can be seen from Fig.~\ref{fig:shifts}a. For low values of $a/\delta$ we have $-\Delta f/(\alpha f_0)\!\sim \!(a/\delta)^2\!\sim\!\omega$, the same as predicted by the Sauerbrey Eq.~(\ref{sauerbrey}) (see Fig.~\ref{fig:shifts}a), while similarly  $\Delta\Gamma/(\alpha f_0)\!\sim\! \omega^{3/2}$. At higher values of $a/\delta \gtrsim \! 0.2$ the crossover to a sublinear dependence $\Delta\Gamma/(\alpha f_0)\!\sim\! \omega^{0.82}$ can be observed, and similarly the crossover to $\Delta\Gamma/(\alpha f_0)\!\sim\!\omega^{0.64}$ occurs for $a/\delta\! \gtrsim \! 0.5 $. As a result, the dimensionless acoustic ratio, $\Delta\Gamma/(-\Delta f)$, shows a non-monotonic dependence on $a/\delta$, reaching the maximum value at $a/\delta\!\approx \! 0.8$--$0.9$ depending on $\rho_s$. (see Fig.~\ref{fig:shifts}b).

The idea to use QCM-D as a tool for quantifying the rheological (viscoelastic) properties of the contact sounds appealing, but its implementation is perhaps too complex. The accurate account of hydrodynamics in such case would be difficult, as, subtle differences in particle mobility produces large differences in the excess shear force, e.g., notice the difference in impedance due to freely suspended (near contact, see \cite{prf23}) and rigidly attached particles. Moreover, for compliant contact yields the particle's motion with respect to the resonator (i.e., rocking or sliding \cite{johan09}) which is determined by the interplay of adhesive and hydrodynamic forces, rendering the accurate quantitative analysis of the QCM-D signal extremely difficult. 

Apparently in the experiments the contact elasticity and/or finite particle deformability affects the signal at higher overtones, suggesting that perhaps accurate gravimetric measurements are possible at lower frequencies. Since fundamental frequency of AT cut quartz is inversely proportional to its thickness, it is theoretically possible to build a device with a thicker crystal operating at lower resonant frequency. 
Such modification would only be useful provided that at lower frequencies the adhesive contact remains stiff. 

The authors would like to thank D.~Johannsmann, R.~Buscalioni and Z. Adamczyk for fruitful discussions. 
This work was supported, in part, by the Israel Science Foundation (ISF) via the grant No. 2899/21. A.M.L. also acknowledges the support of the David T. Siegel Chair in Fluid Mechanics.



\begin{thebibliography}{99}
\bibitem{note0} This is different in the unbounded fluid, where $\mathcal{O}(1)$ velocity at the particle surface produces linear in $\lambda$ terms in the solution of Eqs.~\ref{fdon}. However, in the semi-infinite domain bounded by the no-slip wall at $z=0$, the unsteady term produdes terms of lower order, see the detailed discussion in \cite{fl18}.
\bibitem{note1} Notice that in \cite{oneill68} the torque is showing as $L_y\!=\!-8\pi\mu u a^2 g$, while it should be $-4\pi\mu u a^2 g$; the correct coefficient ($-4\pi$) is provided in Ref.~\cite{brenner67}.
\end{thebibliography}
\end{document}